\documentclass[twocolumn]{aastex62}
\usepackage{graphics,epsf}
\usepackage{amsmath}                
\usepackage{amsfonts}               
\usepackage{amssymb}                
\usepackage{epsfig}                 
\usepackage{graphicx}
\usepackage{float}
\usepackage{color}
\usepackage[para,online,flushleft]{threeparttable}

\newcommand{\cm}{{~\rm cm}}
\newcommand{\km}{{~\rm km}}
\newcommand{\s}{{~\rm s}}

\newcommand{\K}{{~\rm K}}
\newcommand{\erg}{{~\rm erg}}

\newcommand{\kpc}{{~\rm kpc}}

\newcommand{\AU}{{~\rm AU}}

\newcommand{\days}{{~\rm days}}

\begin{document}

\title{Efficiently jet-powered radiation in intermediate-luminosity optical transients (ILOTS)}

\author[0000-0003-0375-8987]{Noam Soker}
\affiliation{Department of Physics, Technion, Haifa, 3200003, Israel;  soker@physics.technion.ac.il}
\affiliation{Guangdong Technion Israel Institute of Technology, Shantou 515069, Guangdong Province, China}

\begin{abstract}
I show that a flow structure where wide jets hit a slower expanding shell might be very efficient in channelling the kinetic energy of the jets to radiation, therefore accounting for, at a least a fraction of, intermediate luminosity optical transients (ILOTs) where the total radiation energy is much larger than what recombination energy of the outflow can supply.
This type of flow might occur in the frame of the high-accretion-powered ILOT (HAPI) model, where there is a high mass accretion rate as a result of stellar merger or mass transfer in a binary system. I derive the condition on the jets half-opening angle for the jets not to penetrate through the slow shell, as well as the ratio of the photon diffusion time to expansion time. This ratio cannot be too large if a large fraction of the thermal energy is channelled to radiation. I apply the jet-powered radiation model to the Great Eruption of Eta Carinae, to V838~Mon, and to V4332~Sgr, and find a plausible set of parameters for these ILOTs. I expect the jet-powered radiation model to be more efficient in converting kinetic energy to radiation than ILOT models that are based on equatorial mass concentration. In many cases, thought, I expect both jets and equatorial mass concentration to occur in the same system. 
\end{abstract}

\keywords{binaries: close --- stars: jets --- stars: variables: general}

\section{Introduction}
\label{sec:intro}

Observations show that eruptive transients with peak luminosities between luminosities of classical novae and typical luminosities of supernovae form a heterogeneous group (e.g. \citealt{Mouldetal1990, Rau2007, Ofek2008, Masonetal2010, Kasliwal2011, Tylendaetal2013, Kasliwal2013, Blagorodnovaetal2017, Kaminskietal2018, Pastorelloetal2018, BoianGroh2019, Caietal2019, Jencsonetal2019, PastorelloMasonetal2019}). These transients might differ from each other by several properties, including by the powering mechanism, which might be thermonuclear or gravitational. The present study aims at eruptions that are powered by gravitational energy. 

The gravitational energy source is a high mass accretion rate, either by a mass transfer or by the destruction of one star onto the other, including a merger and the onset of a common envelope phase (e.g., \citealt{RetterMarom2003, Tylendaetal2011, Ivanovaetal2013a, Nandezetal2014, Kaminskietal2015, MacLeodetal2017, Segevetal2019, Schrderetal2020, MacLeodLoeb2020}) and the  grazing envelope evolution \citep{Soker2016GEEI}. In the case of mass transfer without a merger process, the ILOT might repeat itself, even many times. An intermediate case is an evolution where the secondary star gets in and out of the primary envelope, such as in the common envelope jets supernova (CEJSN) impostor scenario where a neutron star gets in and out from a giant envelope as it launches jets \citep{SokerGilkis2018, Gilkisetal2019, YalinewichMatzner2019}. 

The model of gravitational powering mechanism is termed the high-accretion-powered ILOT (HAPI) model \citep{KashiSoker2016, SokerKashi2016TwoI}. The accretion of mass through an accretion disk might lead to the launching of jets. The focus of this study is the efficient channelling of the kinetic energy of the jets to radiation. 
I use the term Intermediate Luminosity Optical Transients (ILOTs; \citealt{Berger2009, KashiSoker2016, MuthukrishnaetalM2019}) for these gravitationally powered transients. One can define several sub-classes of the heterogeneous class of ILOTs (e.g., \citealt{KashiSoker2016}\footnote{See \url{http://physics.technion.ac.il/~ILOT/} for an updated list.}).
I do note that there is no agreement yet on the division to sub-classes and on what terms to use for the different sub-classes.  \cite{PastorelloMasonetal2019} and \cite{PastorelloFraser2019}, for example, refer to the objects in the relevant luminosity range as gap transients, and refer to those that are powered by binary interaction as luminous red novae. 
They have a different  division of ILOTs to sub-classes than that of \cite{KashiSoker2016}.  Some others also do not use the term ILOT (e.g., \citealt{Jencsonetal2019}). 

I continue with the view of previous studies that attribute ILOT events to binary interaction (e.g., \citealt{Kashietal2010, McleySoker2014, Pejchaetal2016a, Pejchaetal2016b, Soker2016GEEI, MacLeodetal2018, Michaelisetal2018, PastorelloMasonetal2019}).
Particularly, this study deals with processes in the frame of the HAPI model, where the accretion process leads to the launching of jets. The jets collide with a slower outflow and transfer kinetic energy to thermal energy (section  \ref{sec:Thermal}). The conditions on the jets to efficiently thermalize their kinetic energy is that the jets do not penetrate the slowly expanding gas (section \ref{sec:widejets}). 
In section \ref{sec:emission} I study the conditions to channel a large fraction of the thermal energy to radiation. In section \ref{sec:TwoCases} I apply the conclusions to three ILOTs. 

Another way to transfer kinetic energy to radiation is the collision of a spherical fast ejecta with an equatorial disk or an equatorial outflow (e.g., \citealt{AndrewsSmith2018, KurfurstKrticka2019}), where the a binary interaction forms the equatorial mass concentration  (e.g., \citealt{Pejchaetal2016a, Pejchaetal2016b, AndrewsSmith2018, HubovaPejcha2019, KurfurstKrticka2019, GofmanSoker2019}). I compare these cases to jet-powered radiation ILOTs in section \ref{sec:Comparison}. 
I summarise this study in section \ref{sec:summary}. 

\section{Large thermal energy}
\label{sec:Thermal}

I consider cases where the interaction is very strong in the sense that (1) a large fraction (but not necessarily the majority) of the kinetic energy of the outflow is channelled to radiation, and (2) the total radiated energy is larger than what the recombination energy can supply.

The first condition requires a strong collision between different ejecta components, like the collision of two or more shells, jets with a shell, jets with blobs, etc. In this study I take the slow outflow to be a spherical shell of mass $M_{\rm s}$, and a velocity of $v_{\rm s}$. {{{{ Such a shell might be a consequence of the strong gravitational interaction of a companion with a giant star (the primary). For example, a companion on an eccentric orbit perturbed the envelope near a periastron passage, and ejects a shell. Then, later in the orbit it launches the jets from its accretion disk that the  accreted gas forms. This process deserves further study.  }}}}
     
\begin{figure}[t]
	\centering
\includegraphics[trim=23cm 10cm 20cm 0cm ,clip, scale=0.12]{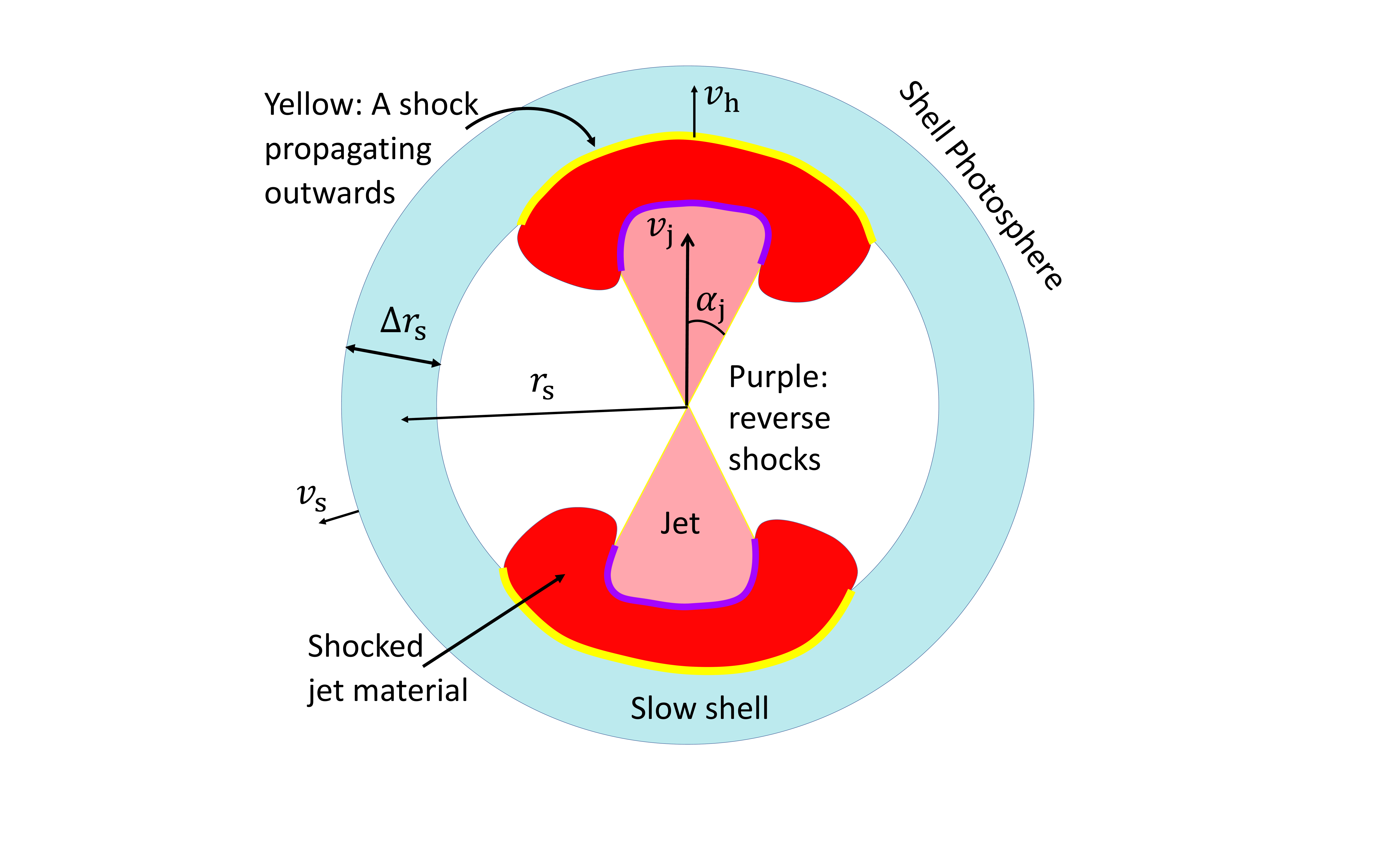}
\caption{A Schematic illustration of the interaction of the jets with the slow shell. The figure does not show the shocked slow shell gas (it is part of the yellow thick line that presents the forward shocks; see \citealt{AkashiSoker2013}). }
	\label{fig:schecmatic}
\end{figure}

For the fast component I take jets, that might be narrow to wide, i.e., from a half opening angle of $\alpha_j \simeq {\rm few}^\circ$ to close to $\alpha_j \la 90^\circ$.  The mass in the two jets is $M_{\rm 2j}$ and they have an initial terminal velocity of $v_{\rm j} \gg v_{\rm s}$. 
{{{{ By `jets' I also refer to a wide biconical outflows, or biconical disk winds. \cite{Bollenetal2019} analyse observations of the post-AGB binary system IRAS19135+3937, and concluded that a stellar jet with a half opening angle of 76 degrees that reaches a velocity of $v_j= 870 \km \s^{-1}$ best fits their observations. They term it a ``jet''despite the very large opening angle. The high velocity, very similar to the escape speed from a main sequence star, suggests that the launching of this very wide jet is from near the main sequence companion, rather than an extended disk-wind. }}}} 
I schematically present the flow in Fig. \ref{fig:schecmatic}. 
{{{{ \cite{AkashiSoker2013} conducted three-dimensional simulations of this type of flow, and the figures there present a better visualisation of the flow interaction. }}}}
   
In this flow most of the kinetic energy is carried by the jets,
$(1/2)M_{\rm 2j} v^2_{\rm s} \gg (1/2) M_{\rm s} v^2_{\rm s}$ (see below). 
The two conditions for a strong shock that transfers a large fraction of the kinetic energy of the jets to thermal energy are 
\begin{equation}
M_{\rm 2j} \la M_{\rm c}; \qquad {\rm and} \qquad v_{\rm j} \gg v_{\rm s}, 
\label{eq:Conditions1}
\end{equation}
where $M_{\rm c} \le M_{\rm s}$ is the mass of the shell with which the jets collide. 
To this we need to add the condition that the hot post-shock gas, of the jets and/or the shell, cooling time due to radiation is not much longer than the cooling time due to adiabatic expansion. This is the subject of section \ref{sec:emission}.  
   
I consider cases where the radiated energy is more than twice the recombination energy. This will require that the kinetic energy of the flow is larger than the recombination energy, as not all the kinetic energy is channelled to radiation. One of the conditions for an efficiently radiating jet-powered ILOT is therefore 
\begin{equation}
E_{\rm rad} > E_{\rm rec} = 3 \times 10^{46}
\left(\frac {M_{\rm s} + M_{\rm 2j}}{M_\odot} \right) \erg,  
\label{eq:Erec}
\end{equation}
assuming that both the hydrogen and helium in the jets and in the shell recombines and the gas has a solar composition. 
For comparison, if a fraction $\eta$ of the kinetic energy of the jets is turned to radiation, then the radiation due to the kinetic energy of the jets is 
\begin{equation}
E_{\rm rad, j}  = 10^{48} \eta  
\left(\frac {M_{\rm 2j}}{0.1 M_\odot} \right) 
\left(\frac {v_{\rm j}}{1000 \km \s^{-1}} \right)^2 \erg .  
\label{eq:Eradj}
\end{equation}

The main conclusion here is that the outflow velocity should be the typical escape speed from main sequence stars or from more compact objects. 

{{{{ The ejection of the shell process itself is accompanied by a bright event. In such a case, there is a possibility to have two peaks in the light curve. In the present study I scale the mass in the jets to be $M_{\rm 2j} \approx 0.1 M_{\rm s}$. Very crudely, I take the radiated energy that accompanied the shell ejection to be about equal to the kinetic energy of the shell, although I expect the radiated energy to be lower even as the shell is very dense when it is ejected (therefore, the photon diffusion time is long; see section \ref{sec:emission}). For the velocity of the shell I take $v_s \approx 10 \km \s^{-1}$ for giant stars of low mass stars and $v_s \approx 100 \km \s^{-1}$ for massive stars, such as Eta Carinae. As well, the ILOT duration time, $\Delta_{\rm ILOT}$, is weeks to months, while the ejection time of the shell, $\Delta t_{\rm s}$, might continue for months to years, i.e.,  $\Delta t_{\rm ILOT} \la \Delta t_{\rm s}$.  Overall, for the ILOTs I examine in this study I expect that the typical luminosity of the shell ejection process, $L_{\rm s}$, be much lower than that due to jets interaction with the shell, $L_{\rm ILOT}$,  
\begin{equation}
\frac{L_{\rm s}}{L_{\rm ILOT}} \la  
0.1 \frac{1}{\eta} 
\left( \frac{M_{\rm s}}{10 M_{\rm 2j}} \right)
\left( \frac {v_{\rm s}}{0.1 v_j} \right)^2  
\left( \frac{\Delta t_{\rm ILOT}}{\Delta t_{\rm s}} \right)  .
\label{eq:LshellLjets}
\end{equation}
But, again, it is possible that there will be a fainter and longer peak in the light curve (an early transient event) that accompanied the shell ejection, therefore occurring months to years before the main ILOT event. }}}}

\section{The requirement for wide jets}
\label{sec:widejets}
 
To utilise a large fraction of the kinetic energy of the jets to radiation, the jets should not penetrate through the slow shell and escape. The jets collide with the slow shell and exert a ram pressure of about $\rho_{\rm j} (v_{\rm j} - v_{\rm h})^2$, where $v_{\rm h}$ is the velocity of the head of the jets, namely, the velocity of the interaction zone between the jet and the shell (yellow boundaries in Fig. \ref{fig:schecmatic}). The slow shell exerts a ram pressure on the interaction zone in the opposite direction $\rho_s (v_{\rm h}-v_{\rm s})^2$, where $\rho_{\rm s}$ is the density in the slow shell. 
Let the head of the jet proceeds through the slow shell with a velocity that obeys $v_{\rm s} \ll v_{\rm h} \ll v_{\rm j}$. Balancing ram pressure on both sides of the interaction zone, $\rho_{\rm j} v^2_{\rm j} \simeq \rho_{\rm s} v^2_{\rm h}$, gives the velocity of the jets' head as 
$v_{\rm h} \simeq v_{\rm j} (\rho_{\rm j} / \rho_{\rm s})^{1/2}$. 
The condition that the jets do not penetrate through the shell reads  
\begin{equation}
\frac{\Delta r_{\rm s}}{v_{\rm h}} \simeq \frac{\Delta r_{\rm s}}{v_{\rm j} (\rho_{\rm j} / \rho_{\rm s})^{1/2}}  > \Delta t_{\rm j},
\label{eq:NonPen1}
\end{equation}
where $r_{\rm s}$ is the average radius of the  shell, $\Delta r_{\rm s}$ is the thickness of the slow shell, and $\Delta t_{\rm j}$ is the jets' launching time period.

The density of the pre-shock jet is 
\begin{equation}
\rho_{\rm j} = \frac {\dot M_{\rm 2j}}{\Omega_{\rm 2j} r^2_s v_{\rm j}}, 
\label{eq:rhoj}
\end{equation}
where $\Omega_{\rm 2j}$ is the solid angle that the two jets span, and 
$\dot M_{\rm 2j}=M_{\rm 2j}/\Delta t_{\rm j}$. Substituting for the density of the jets from equation (\ref{eq:rhoj}) and for the density of the slow shell $\rho_{\rm s} = M_{\rm s}/(4 \pi r^2_s \Delta r_{\rm s})$ in equation (\ref{eq:NonPen1}), we derive the non-penetration condition as 
\begin{eqnarray}
\begin{aligned} 
\frac{\Omega_{\rm 2j}}{4 \pi} \ga  
& 
\frac{M_{\rm 2j}}{M_{\rm s}} 
\frac {v_{\rm j} \Delta t_{\rm j}}{\Delta r_{\rm s}}  
=  0.29 
\left( \frac{M_{\rm 2j}}{0.1 M_{\rm s}} \right)
\left( \frac {v_{\rm j}}{1000 \km\s^{-1}} \right)  
\\ & \times 
\left( \frac{\Delta t_{\rm j}}{10 \days} \right) 
\left( \frac{\Delta r_{\rm s}}{3 \times 10^{13} \cm} \right)^{-1}  .
\end{aligned}
\label{eq:NonPen2}
\end{eqnarray}
This value of $\Omega_{\rm 2j}$ corresponds to a jet's half opening angle of 
$\alpha_{\rm j}=45^\circ$.
 
Different ILOTs will have very different properties, and it might be that in some cases the jets might be narrow. However, equation (\ref{eq:NonPen2}) suggests that in many cases efficiently radiating jet-powered ILOTs require wide jets. 
 
In some cases a close binary system itself forms wide jets even if the outflow from the secondary star that launches the jets is narrow. Consider two very close main sequence stars, or slightly evolved off the main sequence stars. The orbital velocity of the stars is about the Keplerian velocity on the surface of a main sequence star, $\simeq v_{\rm Kep}$. The star that launches the jets orbit the center of mass with a velocity that is a large fraction of that, say $v_{\rm orb} \simeq 0.5 v_{\rm Kep}$. The star launches the jets with the escape velocity from a main sequence star $v_{\rm j} \simeq 2^{1/2} v_{\rm Kep}$. 
Over one orbital period the polar outflow will have a half opening angle that is $> \tan^{-1} (v_{\rm orb} / v_{\rm j}) \simeq \tan^{-1} 0.35 = 19^\circ$, for these values.   

The requirement by equation (\ref{eq:NonPen2}) implies that in many cases of efficiently radiating jet-powered ILOTs the outflow morphology will be of a bipolar structure with two large opposite lobes, e.g., like the the structure of the bipolar outflow of Eta Carinae, the so called Homunculus. 

\section{Emission efficiency}
\label{sec:emission}
  
The shocked jets material forms a hot zone. For jets that do not penetrate deep into the slow shell this hot gas fills a large  volume within the slow shell \citep{AkashiSoker2013}. This hot gas accelerates the shell and cools by this $P dV$ work. For an efficient conversion of thermal energy to radiation, the photon diffusion time must be shorter, or not too much longer, that the expansion time. {{{{ I repeat here a similar calculation to that of \cite{AkashiSoker2013}. They derived the ratio of photon diffusion time to adiabatic cooling time to explore the regime where jets-shell interaction does not suffer much cooling. They simulated some cases of such interactions. They showed that the interaction is prone to Rayleigh–Taylor instabilities that make the real situation more complicated than the ideal case that I explore below. This section will serve me in studying three specific ILOTs (section \ref{sec:TwoCases}), and then in comparing to other types of interactions (section \ref{sec:Comparison}).   }}}} 

If radiation is not efficient, then energy conservation gives the final velocity of the shell as $v_{\rm s,f} \simeq [(M_{\rm 2j}/(M_{\rm 2j}+M_{\rm s})]^{1/2} v_{\rm j}$. For a jets' mass that is about $0.1$ times the shell mass, for example, $v_{\rm s,f} \simeq 0.3 v_{\rm j}$. The expansion time of the shell, $t_{\rm exp} \approx r/(0.5 v_{\rm s,f})$ is  
\begin{eqnarray}
\begin{aligned} 
t_{\rm exp} \approx & 73    
\left( \frac{ r_{\rm s}}{10^{14} \cm} \right)  
\left( \frac{v_{\rm j}}{1000 \km \s^{-1}} \right)^{-1} 
\\ & \times 
\left[ \frac{M_{\rm 2j}}{0.1(M_{\rm 2j}+M_{\rm s})} \right]^{-1/2}
\days.  
\end{aligned}
\label{eq:texp}
\end{eqnarray}

The photon diffusion time across the shell is 
\begin{eqnarray}
\begin{aligned} 
t_{\rm diff} & \simeq \frac{3 \tau \Delta r_{\rm s}}{c} 
\simeq  55   
\left( \frac{M_{\rm s}}{1 M_\odot} \right)
\left( \frac{\kappa}{0.1} \right) 
\\ & \times 
\left( \frac{r_{\rm s}}{10^{14} \cm} \right)^{-1}
\left( \frac{\Delta r_{\rm s}}{0.3 r_{\rm s}} \right) \days,
\end{aligned}
\label{eq:tdiff}
\end{eqnarray}
where $\tau=\rho_{\rm s} \kappa \Delta r_{\rm s}$ is the optical depth of the shell, $\kappa$ is the opacity, and $c$ is the light speed.  

The condition that the shocked jets' material cools mainly by radiation reads therefore 
\begin{eqnarray}
\begin{aligned} 
1 & \ga \frac { t_{\rm diff}}{t_{\rm exp}} \approx 0.75  %
\left( \frac{M_{\rm s}}{1 M_\odot} \right)
\left( \frac{\kappa}{0.1} \right) 
\left( \frac{v_{\rm j}}{1000 \km \s^{-1}} \right)
\\ & \times 
\left( \frac{r_{\rm s}}{10^{14} \cm} \right)^{-2}
\left( \frac{\Delta r_{\rm s}}{ 0.3 r_{\rm s}} \right) 
\left[ \frac{M_{\rm 2j}}{0.1(M_{\rm 2j}+M_{\rm s})} \right]^{1/2}.
\end{aligned}
\label{eq:tdiffdexp}
\end{eqnarray}
The fraction of energy that ends in radiation out of the total energy is $f \simeq t^{-1}_{\rm diff}/(t^{-1}_{\rm diff} + t^{-1}_{\rm exp})$.
The inequality in equation (\ref{eq:tdiffdexp}) holds for cases where more energy ends in radiation than in kinetic energy. But for the cases I study here, even somewhat higher values of $t_{\rm diff}/t_{\rm exp} \la 10$ are fine, implying that about $10 \%$ or more of the energy  ends in radiation. 

{{{{ An interesting property of equation (\ref{eq:tdiffdexp}) is that it does not depend on the opening angle of the jets, as long as it obeys the inequality (\ref{eq:NonPen2}). In particular, it holds for a spherical fast wind instead of jets.  I consider jets for two reasons. (1) The bipolar nebulae around some evolved stars, e.g., the Homunculus of Eta Carinae 
(see section \ref{subsec:EtaCarinae}), suggest the action of jets. (2) For non exploding stars (supernovae), it is not easy to come up with a scenario that has a fast spherical wind that follows a slow wind within a time scale of several years or less (this timescale of several years is the expansion time to a distance of $r_s \approx 10^{14} \cm$). In any case, if there is a scenario for such a rapid transition from a slow wind to a fast spherical wind, then equation (\ref{eq:tdiffdexp}) is applicable. Another condition of course is that the fast wind should be massive enough to carry a large enough kinetic energy to cause a bright event that is observed as an ILOT (equations \ref{eq:Erec}-\ref{eq:LshellLjets}).    
}}}}

\section{Examining three ILOTs}
\label{sec:TwoCases}

I apply the results of the previous sections to the Great Eruption of Eta Carinae, to V838~Mon, and to V4332~Sgr. There are large uncertainties in the values of different parameters, and in the exact morphology of the outflows. Therefore, what I give below are plausible values for these three ILOTs. 

\subsection{The Great Eruption of Eta Carina}
\label{subsec:EtaCarinae}

I consider the powering of the Great Eruption of Eta Carinae, about 1837-1856 
(for a review see, e.g., \citealt{DavidsonHumphreys1997}), by jets that the secondary star launched (e.g., \citealt{Soker2007}) as it accreted mass (e.g., \citealt{KashiSoker2010a,KashiSoker2010b}) from the primary luminous blue variable (LBV) star. 

I take the following general properties of the Great Eruption.
A luminosity of $L_{\rm GE} \simeq 2 \times 10^7 L_\odot$ for a distance of $D=2.3 \kpc$ \citep{DavidsonHumphreys1997}. For the new distance that \cite{Davidsonetal2018} derive, $D=2.6 \kpc$, the luminosity is $L_{\rm GE} \simeq 2.5 \times 10^7 L_\odot$.
The estimates of the effective temperature fall in the range of $T_{\rm eff} \simeq 5000 \K$ \citep{Restetal2012} to $T_{\rm eff} \simeq 6500 \K$ \citep{DavidsonHumphreys2012Natur}. The radius of the photosphere for the new disatnce of $D=2.6 \kpc$ is then $R_{\rm ph} \simeq 4.6 \times 10^{14}-2.7 \times 10^{14} \cm \simeq 31-18 \AU$, for the above range of temperatures, respectively. For the lower distance to Eta Carinae the radii are smaller.  
   
The kinetic energy and mass in the Homunculus (the bipolar outflow) are $E_{\rm kin} \simeq 6 \times 10^{49} \erg$ and $M_{\rm s} \simeq 20 M_\odot$ \citep{Smithetal2003}. 
Adding the extra radiation, $E_{\rm rad,extra} \simeq 2 \times 10^{49}$ (for the old distance of $2.3 \kpc$), the total energy of the event is $\simeq 8 \times 10^{49} \erg$. With the new distance the energy is larger. The observed ratio of radiation to total energy is $f_{\rm GE,o}=[E_{\rm rad} / (E_{\rm kin} + E_{\rm rad})]_{\rm GE,o} \approx 0.25$.  Definitely in the Great Eruption Eta Carinae radiated much more energy than the recombination energy can give (eq. \ref{eq:Erec}).

For the mass that the jets carry I take $M_{\rm 2j} = 2M_\odot$ \citep{KashiSoker2010a} and for their velocity $v_{\rm j} \simeq 3000 \km \s^{-1}$, about the estimated present wind velocity from the secondary star (e.g., \citealt{PittardCorcoran2002}). 

The slow shell might have the escape speed from the primary star, $v_{\rm s} \simeq 500 \km \s^{-1}$. The mass ejected in the event expands relative to the main outflow velocity, with the sound speed, which is $\simeq 10 \km \s^{-1}$. For that, the shell might be thinner even than $\Delta r_{\rm e} \simeq 0.3 r_{\rm s}$, but I will take the value $0.3$. 
Substituting the above values in equation (\ref{eq:tdiffdexp}), I find for the great Eruption of Eta Carina 
\begin{eqnarray}
\begin{aligned} 
\left( \frac { t_{\rm diff}}{t_{\rm exp}} \right)_{\rm GE} \approx  4 &  
\left( \frac{M_{\rm s}}{20 M_\odot} \right)
\left( \frac{\kappa}{0.1} \right) 
\left( \frac{v_{\rm j}}{3000 \km \s^{-1}} \right)
\\ & \times 
\left( \frac{r_{\rm s}}{3.5 \times 10^{14} \cm} \right)^{-2}
\left( \frac{\Delta r_{\rm s}}{ 0.3 r_{\rm s}} \right).
\end{aligned}
\label{eq:tdiffdexpGE}
\end{eqnarray}

The expected ratio of radiated to total energy in this scenario for jet-powered radiation in an ILOT is 
$f_{\rm GE,j} \simeq t^{-1}_{\rm diff}/(t^{-1}_{\rm diff} + t^{-1}_{\rm exp})]_{\rm GE,j} \approx 1/(1+4) \approx 0.2 $. 
Concerning the large uncertainties and some poorly known parameters, this value is quite close to the observed value of $f_{\rm GE,o} \approx 0.25$.  The uncertainties include observational quantities and parameters of the model, like the opacity. In an inhomogeneous (``porous'') atmosphere the effective opacity is lower than that for a homogeneous atmosphere \citep{Shaviv2000}. 
   
The main conclusion of this subsection is that the parameters for the jet-powered radiation model of the Great Eruption of Eta Carinae are plausible. Another support to that model is the bipolar structure of Eta Carinae that has two large opposite lobes with a narrow waist between them. This morphology better fits the expectation of wide jets than that of an efficiently radiating equatorial outflow interaction (section \ref{sec:Comparison}).  
   
\subsection{The ILOT V838 Mon}
\label{subsec:V838Mon}
 
\cite{Tylenda2005} presents a detailed and thorough study of the V838~Mon eruption, and derives the following properties. The effective radius at the end of the eruption was $r \simeq 2 \times 10^{14} \cm$, and the effective temperature during the high luminosity phase was $T_{\rm eff} \simeq 5000-7000 \K$. 
\cite{Tylenda2005} considers a model of shells, but without collision; the first shell in his model is faster than the second shell. The outflow was heated by shocks at the base of the outflow. I expect such a mechanism to be less efficient in channelling energy to radiation than collision between shells (or jets with shell) since the optical depth is large at the inner parts. 
Nonetheless, it is possible that the binary interaction in V838~Mon was not powered by jets at all. {{{{ For example, there are no indications for a bipolar morphology of V838~Mon, unlike the bipolar morphology of the ILOT V4332~Sgr (e.g., \citealt{Kaminskietal2018}). Despite that, }}}}  here I present a plausible set of parameters for jet-powered radiation model of ILOTs with similar radiation properties to those of V838~Mon, even if does not apply to V838~Mon itself.  

\cite{Tylenda2005} estimates the mass of the first shell to be $M_{\rm sh,1} \simeq 0.08 M_\odot$ and its thickness as $0.2$ times its radius, $\Delta r_{\rm sh,1} \simeq 0.2 r_{\rm sh,1}$. For the second shell his estimate is $M_{\rm sh,2} \simeq 0.45-0.65 M_\odot$. \cite{Tylenda2005} claims that if he uses the Planck mean opacity instead of the Rosseland mean opacity, he derives the shell masses to be $M_{\rm sh,1} \simeq 8 \times 10^{-4} M_\odot$ and $M_{\rm sh,2} \simeq 0.004 M_\odot$, respectively.
 
Overall, \cite{Tylenda2005} estimates that V838~Mon ejected a mass in the range of $\approx 0.005M_\odot - 0.6 M_\odot$, \cite{Lynchetal2004} estimate this mass to be $\approx 0.04 M_\odot$, and \cite{TylendaSoker2006AA} suggest a mass loss of $\approx 0.03 M_\odot$ based on an outflow velocity of $\simeq 300 \km \s^{-1}$ and under the assumption that half the energy is carried by radiation and half as kinetic energy.
The total radiated energy in the eruption is $E_{\rm rad,o} \simeq 2.5 \times 10^{46} \erg$ (e.g., \citealt{TylendaSoker2006AA}). 
In that case the total energy of the eruption is $\simeq 5 \times 10^{46} \erg$. But the kinetic energy might be much larger and the total energy of the event might be up to $\simeq 10^{48} \erg$ \citep{Kashietal2010, KashiSoker2010b}. 
      
As a plausible set of values for a jet-powered radiation model in that ILOT I take the following parameters. For the radius I take the value when the rapid photosphere expansion started, which signifies the beginning of the strong interaction between the jets and the slow shell, $r_{\rm s} \simeq 3 \times 10^{13} \cm$. For the jets velocity I take a value of $v_{\rm j} = 700 \km \s^{-1}$, similar to the escape velocity of a main sequence star somewhat more massive that the sun. However, in the present case it is possible that the slow shell that was ejected at the beginning of the interaction is moving at $\simeq 300 \km \s^{-1}$, and has more energy than the jets. The collision then does not transfer all the jets energy to thermal energy, but only a fraction of it, about a half or somewhat more.
Plausible mass values are a total mass in the two jets of $M_{\rm 2j} \la 0.02 M_\odot$, and $M_{\rm s} \simeq 0.2 M_\odot$. For a slow shell velocity of $\simeq 250 \km \s^{-1}$, the total eruption energy is $2.2 \times 10^{47} \erg$. 
Substituting these values in equation (\ref{eq:tdiffdexp}) given 
\begin{eqnarray}
\begin{aligned} 
\left( \frac { t_{\rm diff}}{t_{\rm exp}} \right)_{\rm V838} \approx  1 &  
\left( \frac{M_{\rm s}}{0.2 M_\odot} \right)
\left( \frac{\kappa}{0.1} \right) 
\left( \frac{v_{\rm j}}{700 \km \s^{-1}} \right)
\\ & \times 
\left( \frac{r_{\rm s}}{3 \times 10^{13} \cm} \right)^{-2}
\left( \frac{\Delta r_{\rm s}}{ 0.3 r_{\rm s}} \right) .
\end{aligned}
\label{eq:tdiffdexpV838}
\end{eqnarray}

As said, the shocks convert only a fraction of the kinetic energy, mainly the jets' energy, to thermal energy, $E_{\rm th} \approx 10^{47} \erg$.  
For the values I use in equation (\ref{eq:tdiffdexpV838}), radiation carries about half of the thermal energy $E_{\rm rad} \approx 5\times 10^{46} \erg$. This is about twice the observed value. Considering the large uncertainties, including the value of the opacity \citep{Tylenda2005}, this is an adequate agreement to the purpose of showing that a jet-powered radiation model is a plausible scenario for the eruption of V838~Mon. 

\subsection{The ILOT V4332~Sgr}
\label{subsec:V4332Sgr}

{{{{ I analyse the ILOT V4332~Sgr that has a bipolar structure  \citep{Kaminskietal2018}, which suggests the action of jets. The parameters for this event are less certain than for V838~Mon. \cite{Kaminskietal2018} estimate the ejected mass as $M_{\rm s} \approx 0.01 M_\odot$. The total radiated energy, accounting for several days before first detection, amounts to $\approx 10^{44} \erg$ \citep{Tylendaetal2005}.
\cite{Tylendaetal2005} estimate the radius of the photosphere at peak luminosity to be $\approx 10^{13} \cm$. The interaction must take place at a smaller radius before the photosphere expands, and I take somewhat arbitrarily the radius of the shell to be half that radius, $r_s \approx 5 \times 10^{12} \cm$.   
The velocities from observations are $\simeq 150 \km \s^{-1}$ \citep{Martinietal1999, Kaminskietal2018}; the jets have been faster.
Based on the study of \citep{Tylendaetal2005} the mass in the jets is very small $M_{\rm 2j} \approx 10^{-4} M_\odot \approx 0.01 M_{\rm s}$. For jets' velocity of $v_{\rm j}=700 \km \s^{-1}$ the ratio of radiation to kinetic energy in the jets is 0.2.  In this event, it is likely that the velocity of the slow shell was large, such that the shell caries a large kinetic energy.  }}}}

{{{{  Substituting the above values (with their large uncertainties) in equation (\ref{eq:tdiffdexp}), I find 
\begin{eqnarray}
\begin{aligned} 
& \left( \frac { t_{\rm diff}}{t_{\rm exp}} \right)_{\rm V4332} \approx 0.7  %
\left( \frac{M_{\rm s}}{0.01 M_\odot} \right)
\left( \frac{\kappa}{0.1} \right) 
\left( \frac{v_{\rm j}}{700 \km \s^{-1}} \right)
\\ & \times 
\left( \frac{r_{\rm s}}{5 \times 10^{12} \cm} \right)^{-2}
\left( \frac{\Delta r_{\rm s}}{ 0.3 r_{\rm s}} \right) 
\left[ \frac{M_{\rm 2j}}{0.01(M_{\rm 2j}+M_{\rm s})} \right]^{1/2}.
\end{aligned}
\label{eq:V4332Sgr}
\end{eqnarray} 
}}}} 

{{{{ For these values, the ratio of the energy in radiation to the kinetic energy of the jets is $1.4$, as compared to the value of $0.2$ cited above. However, there are large uncertainties in the values of the radius of interaction, of the opacity, and of the masses in the jets and shell. I consider the jet-powered radiation model plausible for V4332~Sgr.  }}}} 

\section{Polar interaction versus equatorial interaction}
\label{sec:Comparison}
 
From pure flow considerations, the same treatment of the fast polar outflows, i.e., jets, that hit a spherical slow outflow holds for a fast equatorial flow that hits a spherical shell. However, although such equatorial fast outflow studies exit (e.g., \citealt{Soker1999}), most studies of extra emission from equatorial outflows consider an early slow equatorial flow, e.g., through the outer Lagrange point (e.g., \citealt{Pejcha2014}) and a later fast spherical outflow, e.g., a supernovae explosion, that collides with the slow equatorial outflow (e.g., \citealt{AndrewsSmith2018} for iPTF14hls; for simulations of such a flow, see, e.g., \citealt{KurfurstKrticka2019}).

The internal collisions of a spiral structure that mass loss from the outer Lagrange point forms, thermalize  only $\approx 10 \%$ of the kinetic energy (e.g., \citealt{Pejchaetal2016a}). As the terminal velocity of this equatorial outflow is $v_{\rm s,0} \la 0.25 v_{\rm esc,B}$, where $v_{\rm esc,B}$ is the escape velocity from the binary system, and some of the thermal energy is not radiated, the overall radiated energy per unit mass is $e_{\rm rad} \la 0.01 v^2_{\rm esc,B}/2$ (also \citealt{Pejchaetal2016b}). This is too low an efficiency for the systems I consider here.  

The case of a spherical fast ejecta that hits a dense equatorial outflow can be more efficient than internal collision of the equatorial outflow (e.g., \citealt{MetzgerPejcha2017}). The case of a core collapse supernova explosion with a circumstellar disk is such a case (e.g., \citealt{AndrewsSmith2018}). It might in principle work also for classical nova outbursts. However, in the case of main sequence stars or giant stars, it is not easy to come up with a scenario of a fast spherical ejection event that runs into a disk. In any case, the collision of the fast ejecta with the equatorial outflow thermalizes only a fraction of $\simeq h/r$, where $h$ is the half-thickness of the equatorial outflow at distance $r$. \cite{MetzgerPejcha2017} use $h/r=0.3$ in their study, and \cite{GofmanSoker2019} find that they require a value of $h/r \simeq 0.2 $ to explain the luminosity of the supernova iPTF14hls by ejecta-torus interaction.   
    
Overall, the jet-powered radiation model predicts for many cases (but not all) the formation of large lobes with a narrow waist between them. In case of a fast ejecta interacting with a thick torus, the waist is expected to be thicker, and the lobes will be part of a large sphere where the equatorial part is missing. I think the morphology of Eta Carinae nebula better fits shaping by jets. 

\cite{Kashi2010} suggests that the seventeenth century eruption of the LBV P~Cygni was powered by a mass transfer to a B-type binary companion during periastron passages, and that the B-type companion launched jets (for more indications for a possible companion see \citealt{Michaelisetal2018}). \cite{Kashi2010} further notices that the jets might account for the morphology of the nebula that is peculiar (e.g., \citealt{Notaetal1995}), but contains an axisymmetric component \citep{SmithHartigan2006}. However, for this specific nebula a dense equatorial mass loss might also account for the morphology \citep{Notaetal1995}.
     
Another difference between jet-powered ILOTs and the mass loss from the outer Lagrange point is the duration of mass loss relative to the binary orbit. 
The mass loss through the outer Lagrange point that forms an equatorial outflow lasts for many, up to thousands, of orbits of the binary system prior to the final dynamical coalescence. However, in some cases interaction might be much shorter. For example, in the Great Eruption of Eta Carinae the orbit of the binary system is highly eccentric, and the outbursts lasted for about 4 orbital periods. As well, the bipolar structure of the nebula, the Homunculus, points to shaping by jets. The pre-explosion outbursts of SN~2009ip (those of 2009 and 2011; \citealt{Mauerhanetal2013}) also lasted for a time shorter than few times the orbital period of a possible binary system \citep{Kashietal2013}. In these cases a different powering mechanism than mass loss through the outer Lagrange point must take place. 

{{{{ Consider the observed bipolar structures in some nebulae around evolved stars, e.g., the Homunculus of Eta Carinae. In addition to the model for bipolar nebula formation where jets expand into a spherical slow wind, there is the model where a fast spherical wind collides with a slow equatorial outflow or with a bound post common envelope that collimate the wind onto a bipolar outflow (jets; e.g., \citealt{Franketal2018, GarciaSeguraetal2018, MacLeodetal2018, Zouetal2020}). For the purpose of the present study, that scenario requires three types of flow, as follows, from early to late outflow components. ($i$) A slow outflow (the most outer one) with which the collimated bipolar outflow collides; ($ii$) an equatorial dense outflow or envelope that collimates the fast spherical wind; ($iii$) a fast spherical (or wide) wind, the last to be blown, from inside that the equatorial flow collimates and which then collides with the outer slow outflow. The outcome of the collision is similar to the one that I study here (and that requires two flow components). }}}}

Another direct energy source is the collision of the secondary star with the envelope of the giant star. Namely, the companion crosses the envelope (the companion survives the interaction), and a fraction of its kinetic energy is transferred to radiation. The radiation efficiency might be very low because of these reasons. 
If the larger star is a red giant of any sort, then the velocity of the companion is about the Keplerian velocity on the surface of the giant, which is slow for giant stars. In case of two main sequence stars the velocity is high. In both of these cases the interaction takes place inside a dense envelope so that the photon diffusion time is long, and the hot shocked gas expands and cool adiabatically before it radiates much of the energy. The radiation efficiency of such a process might be low.  

\section{Summary}
\label{sec:summary}
   
This study aims at ILOTs (not at supernovae) where the energy source is gravitational energy that results from a mass transfer event in a binary system, or from the merger of the two stars. In particular, I considered cases where the radiation of the ILOT carries a large fraction, $\ga 10 \%$, of the kinetic energy of the outflow, and the radiated energy is much larger than the recombination energy of the  outflowing gas (section \ref{sec:Thermal}). 

I studied some expected properties of a scenario, the jet-powered radiation model, where jets collide with a slow shell (Fig. \ref{fig:schecmatic}). This collision transfers kinetic energy to thermal energy. For this process to be efficient, the jets should be relatively wide to prevent their penetration through the shell (eq. \ref{eq:NonPen2}).
 
To convert a large fraction of the thermal energy to radiation, the photon diffusion time cannot be much longer than the expansion time scale (the adiabatic cooling time scale). In equation (\ref{eq:tdiffdexp}) I give this condition for about half of the thermal energy to be radiated. But here I considered also cases with somewhat lower efficiency, but larger than $10 \%$ efficiency, in converting kinetic energy to radiation.  
  
In section \ref{sec:TwoCases} I applied the results to the Great Eruption of Eta Carinae, to V838~Mon, and to V4332~Sgr. In these cases I showed that I could find a plausible set of parameters for a jet-powered radiation model to explain the radiation in these three ILOTs. I emphasise again that the uncertainties in some parameters and in the geometry are very large, so there is a need for more studies. 
{{{{ I noted that V838~Mon does not show a bipolar morphology that is expected from the action of jets, unlike the ILOT V4332~Sgr \citep{Kaminskietal2018}. }}}}

In section \ref{sec:Comparison} I compared the jet-powered radiation model to models where there is an equatorial mass concentration rather than polar mass concentration as in jets. I argued that the morphology of the nebula of Eta Carinae (the Homunculus) suggests that the radiation of the Great Eruption was powered mainly by jets. Overall, in ILOTs, I expect the jet-powered radiation model to be more efficient in converting kinetic energy to radiation than models that are based on equatorial mass concentration. 
In many cases, thought, I expect both jets and equatorial mass concentration to occur in the same system. 
 
Binary interaction can lead both to the launching of jets and to an equatorial mass concentration, as morphologies of planetary nebulae and of outflows from other evolved binary systems show. Therefore, the main conclusion of this study is that when studying ILOTs, one should be aware both of jets (polar mass concentration) and of equatorial mass concentration.    
 
\section*{Acknowledgements}
    
I thank Amit Kashi, Tomek Kaminski, and an anonymous referee for many valuable comments. This research was supported by a grant from the Israel Science Foundation.

\label{lastpage}

\begin{thebibliography}{}

\bibitem[Akashi \& Soker(2013)]{AkashiSoker2013} {{{{ Akashi, M., \& Soker, N.\ 2013, \mnras, 436, 1961 }}}}


\bibitem[Andrews, \& Smith(2018)]{AndrewsSmith2018} Andrews, J.~E., \& Smith, N.\ 2018, \mnras, 477, 74

\bibitem[Berger et al.(2009)]{Berger2009} Berger, E., Soderberg, A. M., Chevalier, R. A., et al. 2009, \apj, 699, 1850

\bibitem[Blagorodnova et al.(2017)]{Blagorodnovaetal2017} Blagorodnova, N., Kotak, R., Polshaw, J., et al.\ 2017, \apj, 834, 107

\bibitem[Boian, \& Groh(2019)]{BoianGroh2019} Boian, I., \& Groh, J.~H.\ 2019, \aap, 621, A109.

\bibitem[Bollen et al.(2019)]{Bollenetal2019} {{{{ Bollen, D., Kamath, D., Van Winckel, H., et al.\ 2019, \aap, 631, A53 }}}}

\bibitem[Cai et al.(2019)]{Caietal2019} Cai, Y.-Z., Pastorello, A., Fraser, M., et al.\ 2019, \aap, 632, L6


\bibitem[Davidson, \& Humphreys(1997)]{DavidsonHumphreys1997} Davidson, K., \& Humphreys, R.~M.\ 1997, \araa, 35, 1

\bibitem[Davidson et al.(2018)]{Davidsonetal2018} Davidson, K., Helmel, G., \& Humphreys, R.~M.\ 2018, Research Notes of the American Astronomical Society, 2, 133

\bibitem[Davidson, \& Humphreys(2012)]{DavidsonHumphreys2012Natur} Davidson, K., \& Humphreys, R.~M.\ 2012, \nat, 486, E1

\bibitem[Frank et al.(2018)]{Franketal2018} {{{{ Frank, A., Chen, Z., Reichardt, T.,  De Marco, O., Blackman, E., \& Nordhaus, J.\ 2018, Galaxies, 6, 113 }}}}

\bibitem[Garc{\'\i}a-Segura et al.(2018)]{GarciaSeguraetal2018} {{{{ Garc{\'\i}a-Segura, G., Ricker, P.~M., \& Taam, R.~E.\ 2018, \apj, 860, 19 }}}}

\bibitem[Gilkis et al.(2019)]{Gilkisetal2019} Gilkis, A., Soker, N., \& Kashi, A.\ 2019, \mnras, 482, 4233

\bibitem[Gofman, \& Soker(2019)]{GofmanSoker2019} Gofman, R.~A., \& Soker, N.\ 2019, \mnras, 488, 5854

\bibitem[Hubov{\'a}, \& Pejcha(2019)]{HubovaPejcha2019} Hubov{\'a}, D., \& Pejcha, O.\ 2019, \mnras, 489, 891

\bibitem[Ivanova et al.(2013)]{Ivanovaetal2013a} Ivanova, N., Justham, S., Avendano Nandez, J.~L., \& Lombardi, J.~C.\ 2013, Science, 339, 433

\bibitem[Jencson et al.(2019)]{Jencsonetal2019} Jencson, J.~E., Kasliwal, M.~M., Adams, S.~M., et al.\ 2019, \apj, 886, 40

\bibitem[Kami{\'n}ski et al.(2015)]{Kaminskietal2015} Kami{\'n}ski, T., Mason, E., Tylenda, R., \& Schmidt, M.~R.\ 2015, \aap, 580, A34

\bibitem[Kaminski et al.(2018)]{Kaminskietal2018} Kaminski, T., Steffen, W., Tylenda, R.,  Young, K.~H., Patel, N.~A., \& Menten, K.~M.\ 2018, \aap, 617, A129

	
\bibitem[Kasen, \& Woosley(2009)]{KasenWoosley2009} Kasen, D., \& Woosley, S.~E.\ 2009, \apj, 703, 2205.
  
\bibitem[Kashi(2010)]{Kashi2010} Kashi, A.\ 2010, \mnras, 405, 1924
  
\bibitem[Kashi et al.(2010)]{Kashietal2010} Kashi, A., Frankowski, A., \& Soker, N.\ 2010, \apjl, 709, L11

\bibitem[Kashi, \& Soker(2010a)]{KashiSoker2010a} Kashi, A., \& Soker, N.\ 2010a, \apj, 723, 602

\bibitem[Kashi, \& Soker(2010b)]{KashiSoker2010b} Kashi, A., \& Soker, N.\ 2010b,  arXiv:1011.1222

\bibitem[Kashi \& Soker(2016)]{KashiSoker2016} Kashi, A., \& Soker, N.\ 2016, Research in Astronomy and Astrophysics, 16, 99

\bibitem[Kashi et al.(2013)]{Kashietal2013} Kashi, A., Soker, N., \& Moskovitz, N.\ 2013, \mnras, 436, 2484

\bibitem[Kasliwal(2011)]{Kasliwal2011} Kasliwal, M.~M.\ 2011, Bulletin of the Astronomical Society of India, 39, 375

\bibitem[Kasliwal(2013)]{Kasliwal2013} Kasliwal, M.~M.\ 2013, IAU Symposium, 281, 9

\bibitem[Kurf{\"u}rst \& Krti{\v{c}}ka(2019)]{KurfurstKrticka2019} Kurf{\"u}rst, P., \& Krti{\v{c}}ka, J.\ 2019, \aap, 625, A24

\bibitem[Lynch et al.(2004)]{Lynchetal2004} Lynch, D.~K., Rudy, R.~J., Russell, R.~W., et al.\ 2004, \apj, 607, 460

\bibitem[MacLeod \& Loeb(2020)]{MacLeodLoeb2020} {{{{ MacLeod, M., \& Loeb, A. 2020, 	arXiv:2003.01123 }}}}

\bibitem[MacLeod et al.(2017)]{MacLeodetal2017} MacLeod, M., Macias, P., Ramirez-Ruiz, E., Grindlay, J., Batta, A., \& Montes, G.\ 2017, \apj, 835, 282
   
\bibitem[MacLeod et al.(2018)]{MacLeodetal2018} MacLeod, M., Ostriker, E.~C., \& Stone, J.~M.\ 2018, \apj, 868, 136.

\bibitem[Martini et al.(1999)]{Martinietal1999} {{{{ Martini, P., Wagner, R.~M., Tomaney, A., Rich R.~M., della Valle M., \& Hauschildt P.~H.,\ 1999, \aj, 118, 1034 }}}}

\bibitem[Mason et al.(2010)]{Masonetal2010} Mason, E., Diaz, M., Williams, R.~E., Preston, G., \& Bensby, T.\ 2010, \aap, 516, A108

\bibitem[Mauerhan et al.(2013)]{Mauerhanetal2013} Mauerhan, J.~C., Smith, N., Filippenko, A.~V., et al.\ 2013, \mnras, 430, 1801

\bibitem[Mcley \& Soker(2014)]{McleySoker2014} Mcley, L., \& Soker, N.\ 2014, \mnras, 440, 582

\bibitem[Metzger, \& Pejcha(2017)]{MetzgerPejcha2017} Metzger, B.~D., \& Pejcha, O.\ 2017, \mnras, 471, 3200
 
\bibitem[Michaelis et al.(2018)]{Michaelisetal2018} Michaelis, A.~M., Kashi, A., \& Kochiashvili, N.\ 2018, \na, 65, 29

\bibitem[Mould et al.(1990)]{Mouldetal1990} Mould, J., Cohen, J., Graham, J.~R., et al.\ 1990, \apjl, 353, L35

\bibitem[Muthukrishna et al.(2019)]{MuthukrishnaetalM2019} Muthukrishna, D., Narayan, G., Mandel, K.~S., Biswas, R., \& Hlo{\v z}ek, R.\ 2019, \pasp, 131, 118002
  
\bibitem[Nandez et al.(2014)]{Nandezetal2014} Nandez, J.~L.~A., Ivanova, N., \& Lombardi, J.~C., Jr.\ 2014, \apj, 786, 39

\bibitem[Nota et al.(1995)]{Notaetal1995} Nota, A., Livio, M., Clampin, M., et al.\ 1995, \apj, 448, 788

\bibitem[Ofek et al.(2008)]{Ofek2008} Ofek, E.~O., Kulkarni, S.~R., Rau, A., et al.\ 2008, \apj, 674, 447


\bibitem[Pastorello \& Fraser(2019)]{PastorelloFraser2019} Pastorello, A., \& Fraser, M.\ 2019, Nature Astronomy, 3, 676

\bibitem[Pastorello et al.(2018)]{Pastorelloetal2018} Pastorello, A., Kochanek, C.~S., Fraser, M., et al.\ 2018, \mnras, 474, 197

\bibitem[Pastorello et al.(2019)]{PastorelloMasonetal2019} Pastorello, A., Mason, E., Taubenberger, S., et al.\ 2019, \aap, 630, A75

\bibitem[Pejcha(2014)]{Pejcha2014} Pejcha, O.\ 2014, \apj, 788, 22

\bibitem[Pejcha et al.(2016a)]{Pejchaetal2016a} Pejcha, O., Metzger, B.~D., \& Tomida, K.\ 2016a, \mnras, 455, 4351
  
\bibitem[Pejcha et al.(2016b)]{Pejchaetal2016b} Pejcha, O., Metzger, B.~D., \& Tomida, K.\ 2016b, \mnras, 461, 2527

\bibitem[Pittard, \& Corcoran(2002)]{PittardCorcoran2002} Pittard, J.~M., \& Corcoran, M.~F.\ 2002, \aap, 383, 636

\bibitem[Rau et al.(2007)]{Rau2007} Rau, A., Kulkarni, S.~R., Ofek, E.~O., \& Yan, L.\ 2007, \apj, 659, 1536

\bibitem[Rest et al.(2012)]{Restetal2012} Rest, A., Prieto, J.~L., Walborn, N.~R., et al.\ 2012, \nat, 482, 375

\bibitem[Retter \& Marom(2003)]{RetterMarom2003} Retter, A., \& Marom, A.\ 2003, \mnras, 345, L25

\bibitem[Schr{\o}der et al.(2020)]{Schrderetal2020} Schr{\o}der, S.~L., MacLeod, M., Loeb, A., Vigna-G{\'o}mez, A., \& Mandel, I.\ 2020, arXiv:1906.04189

\bibitem[Segev et al.(2019)]{Segevetal2019} Segev, R., Sabach, E., \& Soker, N.\ 2019, \apj, 884, 58

\bibitem[Shaviv(2000)]{Shaviv2000} Shaviv, N.~J.\ 2000, \apjl, 532, L137
  
\bibitem[Smith et al.(2003)]{Smithetal2003} Smith, N., Gehrz, R.~D., Hinz, P.~M., et al.\ 2003, \aj, 125, 1458

\bibitem[Smith, \& Hartigan(2006)]{SmithHartigan2006} Smith, N., \& Hartigan, P.\ 2006, \apj, 638, 1045

\bibitem[Soker(1999)]{Soker1999} Soker, N.\ 1999, \mnras, 303, 611		

\bibitem[Soker(2007)]{Soker2007} Soker, N.\ 2007, \apj, 661, 490

\bibitem[Soker(2016)]{Soker2016GEEI} Soker, N.\ 2016, \na, 47, 16

\bibitem[Soker, \& Gilkis(2018)]{SokerGilkis2018} Soker, N., \& Gilkis, A.\ 2018, \mnras, 475, 1198

\bibitem[Soker, \& Kashi(2016)]{SokerKashi2016TwoI} Soker, N., \& Kashi, A.\ 2016, \mnras, 462, 217

\bibitem[Tylenda(2005)]{Tylenda2005} Tylenda, R.\ 2005, \aap, 436, 1009

\bibitem[Tylenda et al.(2005)]{Tylendaetal2005} {{{{ Tylenda, R., Crause, L.~A., G{\'o}rny, S.~K., \& Schmidt M.~R., \ 2005, \aap, 439, 651 }}}}

\bibitem[Tylenda et al.(2011)]{Tylendaetal2011} Tylenda, R., Hajduk, M., Kami{\'n}ski, T., et al.\ 2011, \aap, 528, A114

\bibitem[Tylenda et al.(2013)]{Tylendaetal2013} Tylenda, R., Kami{\'n}ski, T., Udalski, A., et al.\ 2013, \aap, 555, A16
    
\bibitem[Tylenda, \& Soker(2006)]{TylendaSoker2006AA} Tylenda, R., \& Soker, N.\ 2006, \aap, 451, 223

\bibitem[Yalinewich, \& Matzner(2019)]{YalinewichMatzner2019} Yalinewich, A., \& Matzner, C.~D.\ 2019, \mnras, 490, 312

\bibitem[Zou et al.(2020)]{Zouetal2020} {{{{ Zou, Y., Frank, A., Chen, Z., et al.\ {2020}, arXiv:1912.01647 }}}}

\end{thebibliography}
\end{document}